\documentclass[10pt,a4paper,oneside]{report}
\usepackage{amssymb}
\usepackage{amsmath}
\usepackage{graphicx}
\usepackage{subfigure}
\usepackage{url}
\usepackage{hyperref}
\usepackage{fixltx2e}
\newcommand{\chnoinbook}{5} 
\renewenvironment{abstract}{
\begin{center}
\textbf{Abstract}
\end{center}
\begin{list}{}{
    \setlength{\topsep}{0pt}%
    \setlength{\leftmargin}{1cm}%
    \setlength{\rightmargin}{1cm}%
    \setlength{\listparindent}{\parindent}%
    \setlength{\itemindent}{\parindent}%
    \setlength{\parsep}{\parskip}%
  }%
\item[]}{\end{list}}

\renewcommand{\bibname}{References}
\makeatletter
\renewenvironment{thebibliography}[1]
     {\section*{\bibname}
      \@mkboth{\MakeUppercase\bibname}{\MakeUppercase\bibname}%
      \list{\@biblabel{\@arabic\c@enumiv}}%
           {\settowidth\labelwidth{\@biblabel{#1}}%
            \leftmargin\labelwidth
            \advance\leftmargin\labelsep
            \@openbib@code
            \usecounter{enumiv}%
            \let\p@enumiv\@empty
            \renewcommand\theenumiv{\@arabic\c@enumiv}}%
      \sloppy
      \clubpenalty4000
      \@clubpenalty \clubpenalty
      \widowpenalty4000%
      \sfcode`\.\@m}
     {\def\@noitemerr
       {\@latex@warning{Empty `thebibliography' environment}}%
      \endlist}
\makeatother

\begin{document}
\setcounter{chapter}{\chnoinbook}
\makeatletter \renewcommand{\thefigure}{\chnoinbook.\@arabic\c@figure} \renewcommand{\thetable}{\chnoinbook.\@arabic\c@table} \makeatother
\pagestyle{plain}

\begin{center}
\rmfamily{{\huge Chapter~\chnoinbook. Quantum Cryptography\\}
\bigskip
Dag Roar Hjelme\\
\small Department of Electronics and Telecommunications,\\[-0.4ex]
\small Norwegian University of Science and Technology, NO-7491 Trondheim, Norway\\[-0.4ex]
\small dag.hjelme@iet.ntnu.no\\
\medskip
Lars Lydersen\\
\small Department of Electronics and Telecommunications,\\[-0.4ex]
\small Norwegian University of Science and Technology, NO-7491 Trondheim, Norway\\[-0.4ex]
\small University Graduate Center, NO-2027 Kjeller, Norway\\[-0.4ex]
\small lars.lydersen@gmail.com\\
\medskip
Vadim Makarov\\
\small Department of Electronics and Telecommunications,\\[-0.4ex]
\small Norwegian University of Science and Technology, NO-7491 Trondheim, Norway\\[-0.4ex]
\small University Graduate Center, NO-2027 Kjeller, Norway\\[-0.4ex]
\small makarov@vad1.com\\}
\bigskip
August 7, 2011
\bigskip
\end{center}

\begin{abstract}
This is a chapter on quantum cryptography for the book ``A~Multidisciplinary Introduction to Information Security'' to be published by CRC Press in 2011/2012. The chapter aims to introduce the topic to undergraduate-level and continuing-education students specializing in information and communication technology.
\end{abstract}

\def\chapterfiveimagepath{}

\section{Introduction}

When information is transmitted in microscopic systems, such as single photons (single light particles) or atoms, its information carriers obey quantum rather than classical physics. This offers many new possibilities for information processing, since it is possible to invent novel information processes prevented by classical physics.

Quantum cryptography is the most mature technology in the new field of quantum information processing. Unlike cryptographic techniques where the security is based on unproven mathematical assumptions,\footnote{For instance, the security of RSA public-key cryptography (Chapter~3) rests on the widely-believed assumption that the factorization problem is computationally hard. Although no efficient factorization algorithm is publicly known, it has {\em not} been proven that one does not exist. Shor's algorithm for a {\em quantum computer} already allows efficient factorization, however it remains an open question if and when a scalable quantum computer is built. Furthermore, once a classical encryption is broken, the crack can be applied to today's secrets {\em retroactively.} This is uncomfortable for many types of secret information whose value persists for decades: government and military communication, commercial secrets, as well as certain personal information such as financial and medical records.} the security of quantum cryptography is based on the laws of physics. Today it is developed with an eye towards a future in which cracking of classical public-key ciphers might become practically feasible. For example, a quantum computer might one day be able to crack today's codes. The one-time pad\footnote{It has been proven that a secure cipher needs to use the amount of secret key at least as large as the length of the message \cite{BellSystTechJ-28-656}. The one-time pad (Section 3.2) is one such cipher.} remains unassailable even by such future techniques. The weakness of the one-time pad is that a secret, random, symmetric key as long as the message it is intended to encrypt must be securely distributed to the message's intended receiver. Furthermore, the key can only be used once. Quantum cryptography solves this key distribution problem in a way unfeasible using only classical physics, by exploiting how single quantum particles behave.

The working principles of quantum cryptography can simply be explained by considering information transmission using single photons. A single photon can represent a quantum bit, a so-called {\em qubit.} To determine the qubit value one must measure the representing property of the photon (for example polarization). According to quantum physics, such a measurement will inevitably alter the same property. This is disastrous for anyone trying to eavesdrop on the transmission, since the sender and receiver can easily detect the changes caused by the measurement. Since the security can only be determined after a transmission, this idea can not be used to send the secret message itself. However, it can be used to transmit a secret, random, symmetric key for one-time-pad cryptography. If the transmission is intercepted, the sender and receiver will detect the eavesdropping attempt,\footnote{No such possibility exists if the key is exchanged using classical physics because classical bits can be read, and hence copied without the risk of destroying the original bit value.} the key can be discarded and the sender can transmit another key until a secure key is received.

In spite of the simple principles behind quantum cryptography, the idea was first conceived as late as 1970 in an unpublished manuscript written by Stephen Wiesner. The subject received very little attention until its resurrection by a classic paper published by Charles Bennett and Gilles Brassard in 1984. Currently, the technology required for quantum cryptography is available for real-world system implementations.

The objective of this chapter is to present the working principle of quantum cryptography and to give examples of quantum cryptography protocols and implementations using technology available today. Throughout the chapter we minimize the use of quantum physics formalism and no previous knowledge of quantum physics is required. References are provided for the interested reader who craves for more details. A good starting point is the excellent review by Gisin {\em et al.}\ \cite{RevModPhys-74-145}; also the original paper \cite{JCryptology-5-3} explains the quantum cryptography protocol very well.

\section{Quantum Bit}

All information can be reduced to elementary units, which we call bits. Each bit is a {\em yes} or {\em no} that can be represented by the number 0 or the number 1. However, as we will see, reading and writing this information to a qubit is something quite different from reading and writing this information to a classical bit.

\begin{figure}
\centering
\includegraphics{\chapterfiveimagepath 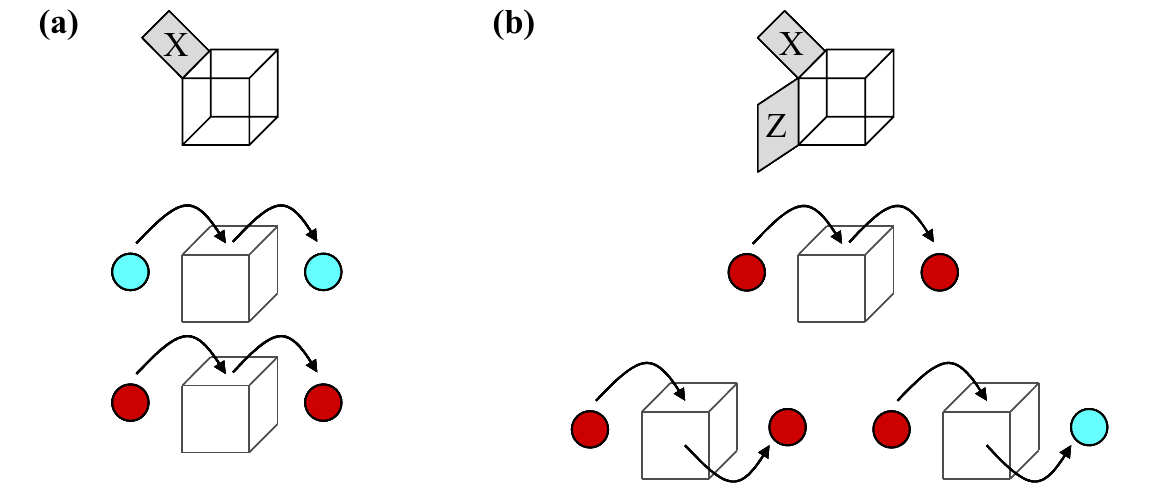}
\caption{Classical versus quantum bit. (a) Classical bit: If we put the ball in a classical box, the color of the ball that pops out is the same as the color we put in. (b) Qubit: If we put the ball in a quantum box and open the wrong door, the color of the ball that comes out is random.}
\label{fig:qcr-qubit-as-a-box}
\end{figure}

We can think of a (qu)bit as a box, where we can store one of the two bit values by putting a ball with one out of two colors into the box as illustrated in Figure~\ref{fig:qcr-qubit-as-a-box}. To read the bit value of the box, we simply open the box and register the color of the ball inside. For the classical bit, the color of the ball inside is always the same as the color of the ball stored in the box in the first place. However, this is not necessarily the case for qubits.\footnote{We have borrowed this way of visualizing a qubit from John Preskill \cite{IEEEAerospaceConfProceedings1998-1-37}.} In the quantum formalism, the two different doors of the box represent two different ways of measuring the qubit value. To read the correct bit information we need to know which door was used when the qubit was stored, and use the same door. If we open the wrong door, the ball inside will have a random color, and thus the information stored in the qubit will change to a random bit value. This also means that the stored information is destroyed.

\begin{figure}
\centering
\includegraphics{\chapterfiveimagepath 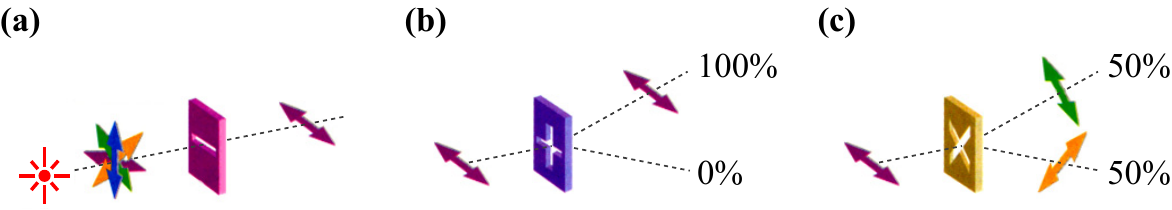}
\caption{Qubit as a polarized photon. (a) A photon source is followed by a linear polarizer to generate a qubit with the desired polarization, in this case a horizontally polarized photon. (b) When a horizontally polarized photon passes through a horizontally-vertically oriented polarizing beamsplitter, it is always found at the exit of the beamsplitter corresponding to the horizontal polarization. (c) When a horizontally polarized photon passes through a diagonally oriented beamsplitter, the photon has 50\% probability to be found at each exit (but the photon will only be detected at one of the exits!). Furthermore, the photon will have a corresponding diagonal polarization afterwards. Therefore, the measurement has changed the state of the photon.}
\label{fig:qcr-qubit-polarization}
\end{figure}

One realization of the qubit is a polarized photon. One way of determining the polarization of the photon is to send it through a polarizing beamsplitter, and measure at which output of the polarizing beamsplitter the photon is found.\footnote{A polarizing beamsplitter is a device that separates orthogonal linear polarizations of incoming light into two directions.} However, since the polarizing beamsplitter only separates between orthogonal polarizations, we cannot orient the polarizing beamsplitter at two angles at the same time. Thus, we can not read the qubit value unless we have additional information. For example, if we know that the polarization is either horizontal or vertical in a defined reference coordinate system\footnote{In quantum physics the orientation of the beamsplitter is called the {\em basis.}} we can read the qubit value by orienting the beamsplitter to the axis of the coordinate system. If we find a photon at one output of the beamsplitter we know that the photon polarization was horizontal; if we find it at another output the polarization was vertical (see Figure~\ref{fig:qcr-qubit-polarization}). That is, we need to know a priori which coordinate system is used in preparing the qubit to read it correctly. If we use another orientation of the beamsplitter, the result of the measurement will be random just like when opening the wrong door of the quantum box in Figure~\ref{fig:qcr-qubit-as-a-box}. Note that once the photon has been detected in one of the outputs after the beamsplitter, the photon actually assumes the output polarization, with no trace of its original polarization left -- this is how the nature works at the quantum level.

\section{Quantum Copying}

To copy a qubit we need to read the bit value, i.e.,\ we need to open the quantum box. However, there is no way of knowing which door was used to store the bit value of the qubit. If we simply guess one of the doors we may damage the information stored in qubit. Thus generally, since quantum bits cannot be perfectly read, quantum bits can not be perfectly copied either.\footnote{For a strict quantum-mechanical proof of this fact, see \cite{Nature-299-802}. The proof is very short.}

Usually, the ability to copy information is considered to be very useful. But, in secure communication, this would be disastrous since the eavesdropper could listen to the communication and keep a copy of the message. However, qubits cannot be copied. This non-copying property of quantum information can be exploited for secure communication. Therefore qubits can be used to distribute a key from sender to recipient without the possibility for the eavesdropper to obtain a copy surreptitiously.

\section{Quantum Key Distribution}

\begin{figure}
\centering
\includegraphics{\chapterfiveimagepath 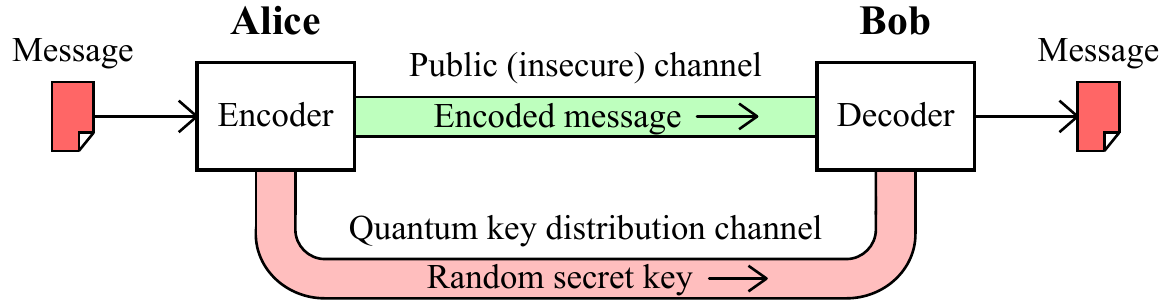}
\caption{Using quantum key distribution in a symmetric encryption scheme. The first step is distribution of a secret key between Alice and Bob. Then, the key can be used by a symmetric cipher to encode and decode transmitted information.}
\label{fig:qcr-key-distribution}
\end{figure}

Quantum cryptography is not used directly to transmit the secret information, but is rather used to distribute a random secret key, see Figure~\ref{fig:qcr-key-distribution}. Once the key has successfully been transmitted, it can be used in a classical symmetric cipher (such as the one-time pad described in Section 3.2 or AES described in Section 2.2.2) to encrypt and decrypt information. Let's consider the quantum key distribution protocol.

\subsection{The BB84 Protocol}

To explain the protocol, let us call the sender Alice, and the receiver Bob. Assume that Alice generates a random sequence of bits, codes them in qubits randomly using door X or door Z of the quantum box, and sends the qubits over a quantum channel to Bob. Bob does not know which doors Alice used, and therefore he randomly picks doors. The result is that Bob opens the right door only half the time. In those cases he reads the right information. Bob's bits are called the raw key at this stage. After Bob has opened all the quantum boxes, both he and Alice publicly announce which doors were used to store and measure the qubits values. They then keep only the qubit values from the boxes where they happened to use the same doors. This random sequence of bits now shared by Alice and Bob is called the sifted key, and is about half as long as the original raw key. 

What happens if the eavesdropper Eve tries to open some of the quantum boxes during the transmission? If Eve by chance opens the right door she can copy the information and send it to Bob. However, half of the time she will open the wrong door and might change the value of the qubit. If Alice and Bob conduct a test and compare a small portion of their key, they can make sure that Bob received what Alice sent. If Alice's and Bob's portion of the key matches, they can be confident that Eve did not open any boxes. On the other hand if their keys do not agree, they know that Eve tried to measure the key.

What we have just described is the quantum key distribution protocol BB84, first presented in 1984 by Bennett and Brassard. Given that Alice and Bob can only measure the fraction of errors in the key, often called the {\em quantum bit error rate,} the protocol either provides a provably secure key or informs Alice and Bob that the key distribution failed.

\subsection{The BB84 Protocol Using Polarized Light}

\begin{figure}
\centering
\includegraphics{\chapterfiveimagepath 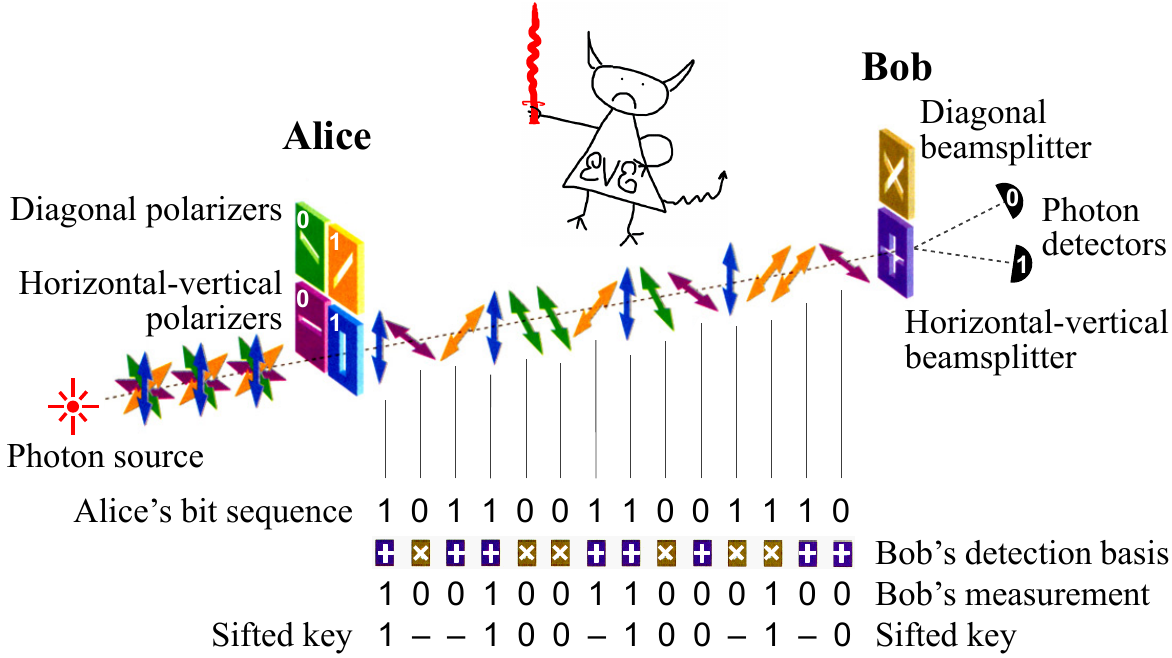}
\caption{BB84 protocol using polarized light (reprinted from \cite{PhysWorld-11-iss3-41}).}
\label{fig:qcr-bb84-principle}
\end{figure}

The BB84 protocol can be implemented using polarized single photons as illustrated in Figure~\ref{fig:qcr-bb84-principle}. Alice codes the qubit using horizontal (bit value 0) and vertical (bit value 1) polarization, or she codes the qubit using $-45^{\circ}$ (bit value 0) or $+45^{\circ}$ (bit value 1) polarization.\footnote{These two ways of doing the coding represent the two doors of the quantum boxes described earlier.} To receive the qubits Bob uses two interchangeable polarizing beamsplitters and two photon detectors\footnote{A photon detector is a device that gives a signal (`click') when a photon arrives at the device.} after the beamsplitter. One polarizing beamsplitter allows Bob to distinguish between the horizontal and vertical polarizations and the other polarizing beamsplitter allows Bob to distinguish between the $-45^{\circ}$ and $+45^{\circ}$ polarizations. If Bob uses a polarizing beamsplitter compatible with the polarization choice of Alice he will read the state of polarization correctly, i.e.,\ he opened the right door. If Bob uses a polarizing beamsplitter incompatible with the polarization choice of Alice he will not be able to get any information about the state of polarization, i.e.,\ he opened the wrong door.

After receiving enough photons, constituting the raw key, Bob announces over a public classical communication channel (e.g.,\ over an internet connection) the sequence of polarizing beamsplitters he used, but importantly not the result of the measurement. Alice compares this sequence to the sequence of bases (polarization choice) she used and tells Bob on which occasions he used the right beamsplitter,\footnote{Effectively publicly announcing which door of the box was used to store each qubit.} but importantly not the polarization she sent. For these bits, constituting the sifted key, Alice and Bob know that they have the same bit values provided that an eavesdropper did not perturb the transmission.

To assess the security of the transmission, Alice and Bob select a random subset of the sifted key and compare it over the public channel. If the transmission were intercepted or perturbed, the correlation between their bit values will be reduced, thus increasing the quantum bit error rate. All eavesdropping strategies perturb the system in some way. Therefore, if Alice and Bob do not measure any discrepancy in the subset of the key, they can be confident that the transmission was not intercepted and they can use the remaining part of the key for encryption.

\section{Practical Quantum Cryptography}

Any implementation of quantum key distribution uses technology available today, meaning that the system components such as photon sources, transmission channel, polarizing beamsplitters, and photon detectors, are imperfect. This fact has several important implications.

One imperfection common to all components is that photons sometimes get lost. In a practical system, the majority of the photons exiting Alice will get absorbed in the transmission channel, and those that reach the detector will often fail to cause a click. In practice only the photons that have registered as clicks in Bob's detectors contribute bits to the raw key.

Another implication of imperfections is that the qubits are prepared and detected not exactly in the basis as described by theory. Technological imperfections will lead to errors in the sifted key, errors that can not be distinguished from errors resulting from any eavesdropping attempts. Realistic error rates with today's technology are in the order of a few percent. This quantum bit error rate is often dominated by false detection signals from the photon detectors, so-called {\em dark counts.}\footnote{Dark counts are clicks in detector without any photons present, and can thus be observed at the detector output in the dark.}

\subsection{Error Correction and Privacy Amplification}

Alice and Bob can not be sure whether the errors in the sifted key resulted from device imperfections or from eavesdropping. They have to assume the worst and assume all errors were due to eavesdropping. At this point in the protocol, Alice and Bob share classical information with high but not 100\% correlation, and assume that the third party Eve has partial knowledge of this information. This problem can be solved by classical information theory, which has methods of distilling a shorter, error-free key of which Eve has no knowledge about.

First, Alice and Bob need to apply classical error correction techniques to obtain identical keys.\footnote{Very high raw error rate of a few percent, while typical for quantum cryptography, usually does not occur in classical telecommunication. Therefore, special error correction algorithms have been developed for quantum cryptography.} Eve still knows some information about this key (actually she knows even more than before, because Alice and Bob have had to reveal more information while communicating publicly during the error correction). The last step in the quantum cryptography protocol therefore is a {\em privacy amplification} procedure that shrinks the key and reduces the amount of information Eve may know about it. Alice and Bob do privacy amplification by applying a randomly chosen hash function of {\em universal}\textsubscript{2}-class to the error-corrected key.\footnote{These hash functions and this application are different from those described in Chapter~4. While the security of cryptographic hash functions in Chapter~4 in not proven, here the security of the privacy amplification procedure \cite{IEEETransInfTheory-41-1915} is {\em unconditional,} i.e.,\ strictly proven against an adversary who possesses unlimited computing power.} As long as Bob has more information about Alice's sifted key than Eve, privacy amplification will produce a shorter final key about which Eve's information is arbitrarily small. To give a feel for the numbers, with realistic quantum bit error rate of 4\%, assumed to be dominated by eavesdropping, 2000 bit can be distilled down to 754 secret bit about which Eve's information is negligible (less than $10^{-6}$ bit). With quantum bit error rate of 8\% we can distill 105 secret bit from the original 2000 bit \cite{JCryptology-5-3}.

The resulting workflow of a general quantum key distribution algorithm is illustrated in Figure~\ref{fig:qcr-post-processing}.

\subsection{Security Proofs}

\begin{figure}
\centering
\includegraphics{\chapterfiveimagepath 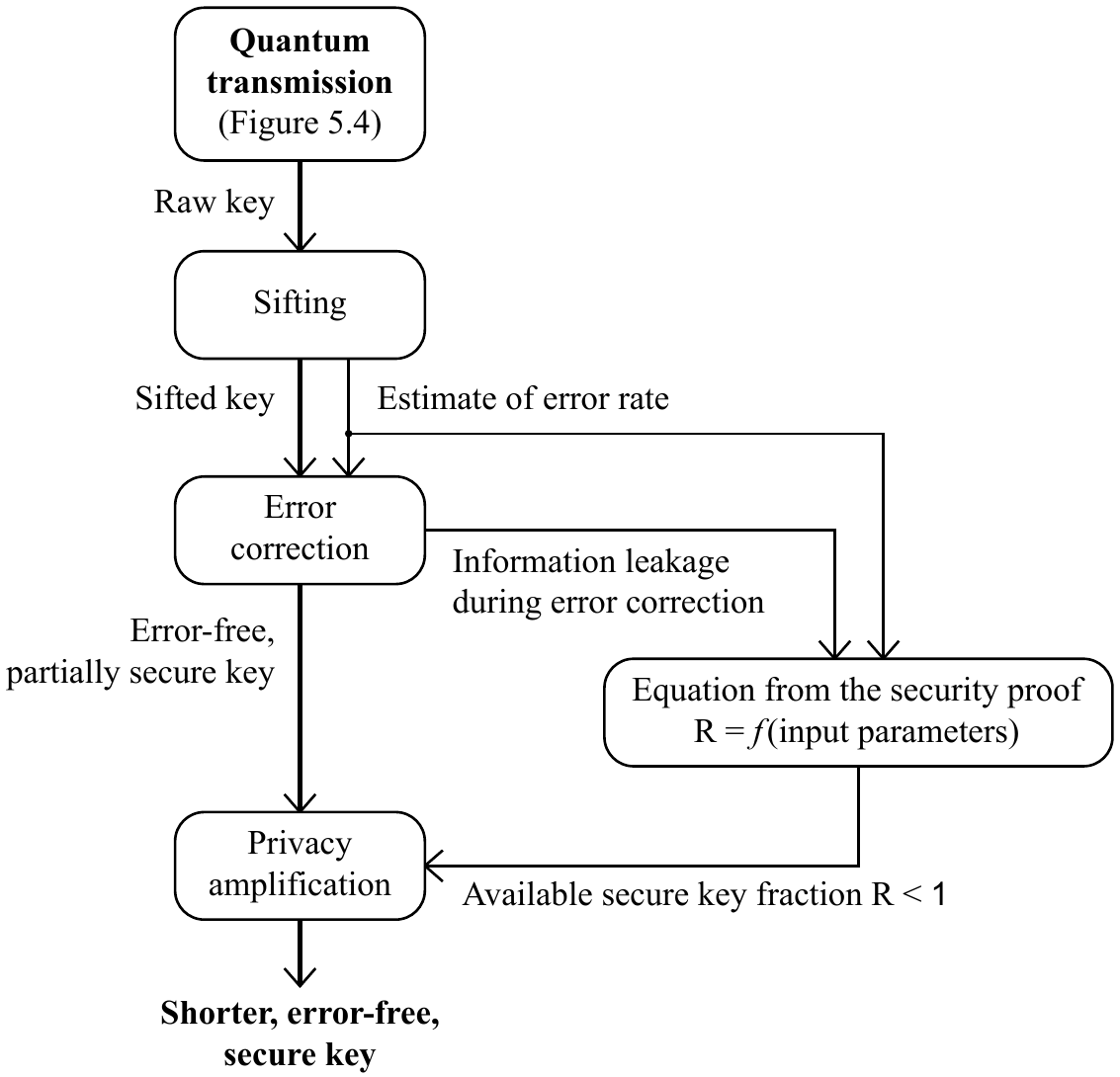}
\caption{Classical post-processing in quantum key distribution. Alice and Bob start with the raw photon detection data, communicate over an authenticated classical channel while performing sifting, error correction and privacy amplification procedures, and arrive at a secret shared key about which Eve has negligible information.}
\label{fig:qcr-post-processing}
\end{figure}

The intuition as to why quantum key distribution provides perfectly secret key is quite straightforward. However, the details of the proofs are very involved \cite{RevModPhys-81-1301}. If one assumes that Eve can only interact with one qubit at a time,\footnote{This is a so-called individual attack.} and that Alice and Bob are using a perfect implementation of the protocol, it has been proven that Eve will never know as much as Bob provided that the quantum bit error rate is less than 14.65\%. If Eve has unlimited power and can coherently attack an unlimited number of qubits,\footnote{This is a so-called coherent attack.} i.e.,\ she can do everything allowed by the known laws of physics, it has been proven that a quantum bit error rate less than 11\% is required for secure communication. As long as the error rate is below this threshold, the security proof provides an equation that can be used to compute the required amount of privacy amplification (Figure~\ref{fig:qcr-post-processing}). 

The security has been proven strictly for certain idealized models of equipment. However, most of the current discussion is whether imperfections in real hardware (not yet accounted by the proofs) may leave loopholes, and how to close these loopholes \cite{NatComm-2-349}.

\subsection{Authentication}

One problem remains: how can Alice and Bob be sure they really talk to each other on the public channel and not to Eve, when they produce secret key? Eve could be in the middle between Alice and Bob, representing herself as Bob to Alice and as Alice to Bob. The prevention of this is known and requires that Alice and Bob start from an initial short common secret (a few hundred bit), so as to be able to recognize each other during their first run of the protocol.\footnote{Unconditionally secure authentication is employed, using hash functions of `almost' {\em universal}\textsubscript{2}-class \cite{JCompSystSci-22-265}. The secret key is used to pick a function from the set of hash functions, then that function is applied to the message to compute a shorter authentication tag. The message and authentication tag are sent to the other communicating party. The latter computes an authentication tag on the received message using her copy of the secret key, and compares it with the received tag. If the tags are the same, this guarantees that the messages are the same with a high probability. The use of one-time pad to pick the hash function guarantees against attacks on this authentication scheme.} After the first successful key distribution, they can use a part of the secret key they produce to authenticate in future runs. It has been proven that quantum key distribution provides much more secret key than it consumes in authentication.\footnote{While perfect encryption, e.g.,\ the one-time pad needs $m$ secret bits to encrypt an $m$-bit message, perfect authentication only needs in the order of $\log(m)$ secret bits to authenticate an $m$-bit message.} In this sense, quantum key distribution is a quantum secret growing protocol.

The need for initial authentication is intrinsic and universal to all flavors of cryptography: how else can you verify that you are talking to the intended party and not to Eve? The initial trusted key and/or biometric authentication (by, e.g.,\ verifying a pen signature, talking to a known person over phone or being physically present during the transaction) is found in some form in all cryptographic protocols.

\section{Technology}

Essentially two technologies make quantum cryptography possible: single photon sources and single photon detectors. In addition, a transmission channel for the single photon states, so-called quantum channel, is needed. The rest of the system is realized using fairly standard telecommunication and electronic hardware.

\subsection{Single Photon Sources}

Single photon sources are difficult to realize. Therefore, most systems today rely on faint laser pulses. Conventional laser pulses, e.g.,\ from a semiconductor laser, are attenuated such that there is on the average less than one photon per pulse. The problem with this approach is that there is a significant probability that there are two or more photons per pulse, unless the average photon number is far below one. The number of photons in the pulse follows Poisson statistics, which for instance means that in a pulse of average photon number 0.1, there is a 0.9048 probability to find no photons, 0.0905 probability to find one photon, and 0.0047 probability to find two or more photons. If Alice emits pulses containing more than one photon, Eve can take and store one of the photons in the pulse until the basis is announced. Then she may perform a perfect measurement in this basis, learning the bit value of the qubit sent to Bob. Therefore, the presence of multiphoton pulses decreases the secret key rate. The fraction of multiphoton pulses relative to single-photon pulses can be reduced by decreasing the average photon number, however when the average photon number is small it means that most bit slots are empty, also resulting in lower bit rate. In principle, the latter could be compensated for by increasing the pulse rate. However, another drawback remains, as the dark counts (false detection events) in the single photon detectors are significant. The result is that the signal to noise ratio decreases, raising the quantum bit error rate, as the average photon number decreases.

The ideal photon source is a device that emits single photons on demand.\footnote{ Such a source is often called a {\em photon gun.}} Although progress is reported, practical devices are not yet available \cite{NaturePhotonics-1-215}.

Nevertheless, practical operation over tens of kilometers has been achieved using faint laser pulse sources. Also, there are advanced protocols\footnote{For example the decoy-state protocol \cite{PhysRevLett-91-057901}.} that allow secure operation over longer than 100~km distance with the faint laser pulse source.

\subsection{Single Photon Detectors}

Single photon detection can be realized in a number of ways, e.g.,\ using photomultipliers, avalanche photodiodes, as well as several types of more exotic superconducting devices that have to be cryogenically cooled below 4~K.\footnote{For a wide review of photon detection techniques, see \cite{NaturePhotonics-3-696}.} Today the best and in fact the only practical choice for quantum cryptography is the avalanche photodiode \cite{JModOpt-51-1267}. An avalanche photodiode is a semiconductor component, and to detect single photons it is operated under a large voltage.\footnote{The photodiode is reverse biased above its breakdown voltage.} If a single photon is absorbed by the semiconductor, it excites a single electron. The high electric field in the semiconductor ensures that this initial electron collides with the lattice and excites more electrons, thus being amplified into an avalanche of electrons (several thousands). This avalanche is large enough to be detected as a current pulse by an external circuit. Unfortunately, an avalanche can also occur without a photon, initiated by thermal excitation, tunneling, or emission of trapped carriers. The latter happens when electrons from a previous avalanche get stuck in defects of the semiconductor lattice, then slowly released. This emission of trapped carriers limits the practical count rate. This is a serious limitation in the current systems using faint laser pulses, where a high pulse rate is desirable in order to achieve acceptable bit rates.

\subsection{Quantum Channel}

Alice and Bob must be connected by a quantum channel. This channel must be such that the qubit is protected from environmental noise. Standard single-mode optical fiber used for data and telecommunication is an almost ideal channel for single photon states (qubits). All optical fibers have transmission losses limiting the number of qubits arriving at the detector. This has direct impact on the key exchange rate, as the raw key rate is directly proportional to the photon transmission probability of the link. Modern telecommunication fibers have transmission losses of about 2~dB/km, 0.35~dB/km and 0.2~dB/km in the commonly used communication wavelength windows of 800~nm, 1300~nm and 1550~nm respectively. At 1550~nm this means that at least 50\% of the photons are lost at 15~km, or 99\% of the photons are lost at 100~km. The longest successful quantum key distribution reported in laboratory conditions to date is 250~km, at a very slow rate of 15 secret bit/s indeed \cite{NewJPhys-11-075003}. Today's commercial systems are limited to 50--100~km.

All fibers are subject to environmental fluctuations, such as a change in temperature. This perturbs a polarization state, and therefore changes the qubit values. Thus, the error rate is increased by the imperfect channel. The global effect of this polarization state perturbation is a transformation between the fiber input and fiber output. If this transformation is stable, Alice and Bob can compensate for it by using a polarization controller to align their systems by defining, e.g.,\ the vertical and diagonal polarization direction. If the transformation varies slowly, one can use an active feedback system to maintain alignment over time. Smart solutions are possible: early commercial systems used a so-called ``plug and play'' optical scheme that cancelled polarization perturbation without a need for active control \cite{ApplPhysLett-70-793}.

As an alternative to fiber, a line-of-sight path via atmosphere can be used as the quantum channel. Alice and Bob use small telescopes pointed at one another to transmit photons. Availability and quality of such a link is obviously affected by weather conditions. However, air neither perturbs polarization nor has a high loss. The longest transmission has been achieved over 144~km between hilltops on the Canary islands \cite{NaturePhysics-3-481}; however links of 10--30~km may be more practical \cite{Nature-419-p450}. The success of these ground experiments suggests a possibility of distributing a secret key between a ground station and a satellite. A low-orbit satellite can thus provide a global key distribution network by successively establishing key distribution links with places under its flight path.\footnote{This requires the satellite to operate as a trusted node, as discussed later in this chapter.}

\subsection{Random Number Generator}

The key used for the one-time pad must be perfectly random. As computers are deterministic systems, they can not be used to create random numbers for cryptographic systems. Therefore, the random numbers must be created by a truly random physical process. One example is a single photon sent through a beamsplitter: the photon is found in one of the two exits of the beamsplitter. Which exit it is found at, is random according to quantum mechanics. There are certainly many other processes that can be used. While physical random number generators with bitrates of a few Mbit/s are employed in current commercial quantum key distribution systems, construction of high bit rate true random number generators is still at an experimental stage.

\section{Applications}

Although quantum cryptography is quite technologically mature, and commercially it currently enjoys a tiny niche market, perspectives of wider adoption are unclear. On the one hand, classical cryptographic systems based on assumptions on computational complexity are very good and convenient: well-developed, cheap, can work at high bitrates over unlimited distance. As we have discussed in the Introduction of this chapter, their security is not guaranteed against future advances in cryptanalysis (and strictly speaking not even guaranteed today), but their convenience is almost unbeatable. Should any of them fall, this would not be the first historical example when the ease of use was preferred over stricter security \cite{Singh1999}.

On the other hand, a quantum key distribution link is limited by distance and bitrate, and is currently relatively expensive to set up. In fact, sending a person (trusted courier) carrying a hard disk filled with random numbers would provide a larger key supply than most quantum key distribution links could deliver over their operation lifetime. Note that the quantum key distribution link also needs a short key for the initial authentication, so the trusted courier is involved anyway. However, the ability to grow key limitlessly from the short authentication key makes quantum cryptography scale well in a network of many users, while in the hard disk distribution scenario the required storage capacity would quickly become unrealistic \cite{Lydersen2011}.

\subsection{Commercial Systems with Dual Key Agreement}

\begin{figure}
\centering
\includegraphics[width=\textwidth]{\chapterfiveimagepath 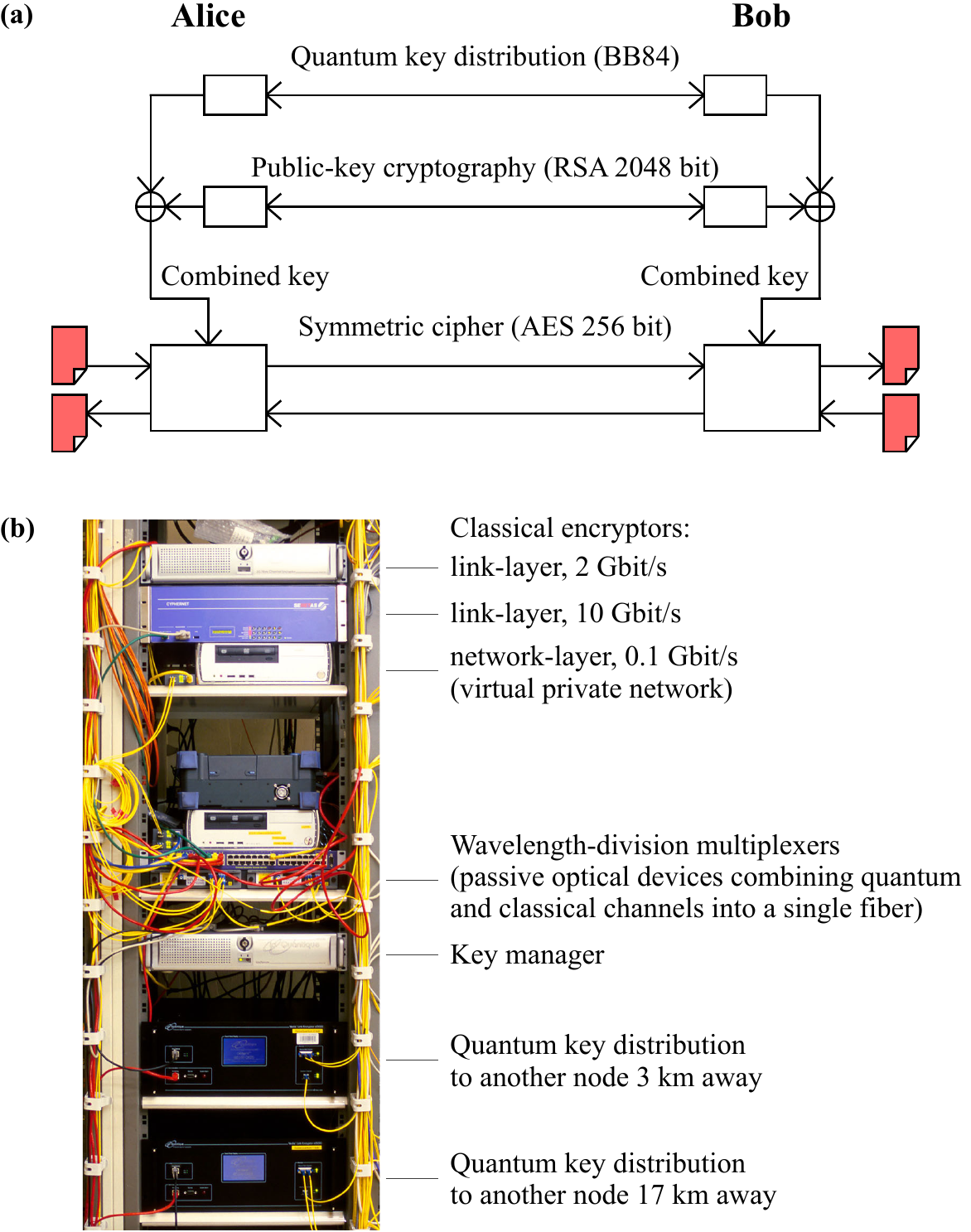}
\caption{Commercial quantum cryptography vintage 2010. (a) Dual key agreement scheme. Two secret keys are distributed independently using quantum cryptography and public-key cryptography, then added modulo~2 (X-ored) together. The resulting key is used in a symmetric cipher to encrypt network traffic. (b) Network node with quantum key distribution equipment, in a standard 19 inch wide server rack. Quantum keys are generated between this node and two other remote nodes using ID~Quantique Vectis units, then passed to classical equipment that encrypts all network traffic with those remote nodes. This node was a part of SwissQuantum testbed network in Geneva, and operated continuously for more than a year, see \url{http://swissquantum.idquantique.com/}.}
\label{fig:qcr-commercial-qcrypto}
\end{figure}

In today's commercial systems, quantum key distribution is used as an extra security layer on top of classical key distribution and encryption, see Figure~\ref{fig:qcr-commercial-qcrypto}. Keys obtained from quantum key distribution are combined with keys sent using public key cryptography, by encrypting one key using the other key as one-time pad (exclusive-OR binary function). The resulting combined key is at least as secure as the stronger of the two original keys. Thus, to eavesdrop the combined key, an attacker would have to crack both public-key cryptography {\em and} quantum key distribution. This combined key is changed several times a second, and used in a high-throughput symmetric cipher to encrypt a wideband network link.

Although any symmetric cipher using key shorter than the encrypted message is not unconditionally secure, this architecture is dictated by the ease of integration into existing networks. Customers are used to having classical cryptography that can encrypt the entire gigabit network link. Nevertheless, it is argued that the security of the AES symmetric cipher improves when the key is changed frequently and thus less ciphertext is available for cryptanalysis.

The system has an option to additionally provide one-time-pad encryption to the users. In the commercially available units, the key generation speed and thus one-time pad average bandwidth is no faster than a few kbit/s, however laboratory prototypes have been demonstrated with up to 1~Mbit/s over 50~km fiber \cite{ApplPhysLett-96-161102}.

\subsection{Quantum Key Distribution Networks}

To increase the number of users and overcome the link distance limitation, two types of networks are possible: with trusted nodes, and with untrusted nodes. The trusted-node network consists of point-to-point quantum key distribution links between nodes. When two users want to establish a shared key, they find a path through intermediate nodes, then one user sends his key to the other user through a chain of one-time-pad encryptions using keys generated in each point-to-point link along the path. This type of network has been demonstrated in several metropolitan areas \cite{NatPhotonics-5-10,NewJPhys-11-075001}.

The untrusted-node network can use optical switches at the nodes to create an uninterrupted optical channel between end users. This is realistic with today's technology, but the optical switches do not increase transmission distance and can thus only be used in a geographically compact network. An alternative idea is to use so-called {\em quantum repeaters} at the untrusted nodes, which in theory can increase the distance far beyond the 250~km limit. However, quantum repeaters remain a future technology. The untrusted-node network configuration can realize the full potential of quantum cryptography, and perhaps provide a decisive advantage over using trusted couriers and other key distribution methods. For example, each user can get and store only initial authentication keys for every other network user, then grow more key material with any user as needed.

\section{Summary}

The feasibility of quantum cryptography has now been demonstrated over distances up to 250~km, and in key distribution networks. Although the systems still suffer from low key transmission rates, they do provide means for secure communication if the public-key systems used today are not trusted. But foremost, today quantum cryptography is developed with an eye towards a future in which cracking classical public-key ciphers might become practically feasible. For example, a quantum computer might one day be able to crack today's codes. Quantum cryptography is also an excellent example of the intimate interplay between fundamental and applied research.

\section{Further Reading and Web Sites}

The web site \url{http://www.iet.ntnu.no/groups/optics/qcr/} of our Quantum Hacking group presents how industrial implementations of the quantum key distribution system can be broken. The web site \url{http://pqcrypto.org/} investigates what will happen to cryptology when the first working quantum computer has been built.

\end{document}